\documentclass[%
reprint,
superscriptaddress,
frontmatterverbose,
showpacs,
preprintnumbers,
nobibnotes,
amsmath,amssymb,
aps,
prc,
]{revtex4-2}
\usepackage[pdftex]{graphicx}
\usepackage{dcolumn}
\usepackage{bm}
\usepackage[pdftex]{color}
\usepackage{CJK}
\usepackage[T1]{fontenc}
\def\ve#1{{\bm{#1}}}
\def\nuc#1#2#3{{}^{#2}_{#3}\mathrm{#1}}
\def\urm#1{\scriptstyle{\text{\textrm{\textmd{\textup{#1}}}}}}

\def\ca#1{{\mathcal{#1}}}
\let\temp\epsilon
\let\epsilon\varepsilon
\let\varepsilon\temp
\let\temp\relax
\let\temp\phi
\let\phi\varphi
\let\varphi\temp
\let\temp\relax
\begin{document}
% 
%%%%%%%%%%%%%%%%%%%%%%%%%%%%%%%%%%%%%%%%%%%%%%%%%% 
\begin{CJK*}{UTF8}{}
  \preprint{RIKEN-QHP-484}
  \preprint{RIKEN-iTHEMS-Report-21}
  \title{
    Toward \textit{ab initio} charge symmetry breaking in nuclear energy density functionals}
  \author{Tomoya Naito (\CJKfamily{min}{内藤智也})}
  \email{
    tomoya.naito@phys.s.u-tokyo.ac.jp}
  \affiliation{
    Department of Physics, Graduate School of Science, The University of Tokyo,
    Tokyo 113-0033, Japan}
  \affiliation{
    RIKEN Nishina Center, Wako 351-0198, Japan}
  \author{Gianluca Col\`{o}}
  \email{
    colo@mi.infn.it}
  \affiliation{
    Dipartimento di Fisica, Universit\`{a} degli Studi di Milano,
    Via Celoria 16, 20133 Milano, Italy}
  \affiliation{
    INFN, Sezione di Milano,
    Via Celoria 16, 20133 Milano, Italy}
  \author{Haozhao Liang (\CJKfamily{gbsn}{梁豪兆})}
  \email{
    haozhao.liang@phys.s.u-tokyo.ac.jp}
  \affiliation{
    Department of Physics, Graduate School of Science, The University of Tokyo,
    Tokyo 113-0033, Japan}
  \affiliation{
    RIKEN Nishina Center, Wako 351-0198, Japan}
  \author{Xavier Roca-Maza}
  \email{
    xavier.roca.maza@mi.infn.it}
  \affiliation{
    Dipartimento di Fisica, Universit\`{a} degli Studi di Milano,
    Via Celoria 16, 20133 Milano, Italy}
  \affiliation{
    INFN, Sezione di Milano,
    Via Celoria 16, 20133 Milano, Italy}
  \author{Hiroyuki Sagawa (\CJKfamily{min}{佐川弘幸})}
  \email{
    sagawa@ribf.riken.jp}
  \affiliation{
    Center for Mathematics and Physics, University of Aizu,
    Aizu-Wakamatsu 965-8560, Japan}
  \affiliation{
    RIKEN Nishina Center, Wako 351-0198, Japan}
  \date{\today}
  %%%%%%%%%%%%%%%%%%%%%%%%%%%%%%%%%%%%%%%%%%%%%%%%%% 
  \begin{abstract}
    We propose a new approach to determine the strength of the charge symmetry breaking (CSB) term in the framework of nuclear density functional theory.
    It is shown that once \textit{ab initio} calculations are available
    including accurate description of isospin symmetry breaking terms in medium and heavy nuclei,
    the mass difference of mirror nuclei
    as well as the neutron-skin thickness of doubly-closed-shell nuclei can be used to constrain 
    the strength of the CSB interaction with an uncertainty less than $ 6 \, \% $,
    separately from other isospin symmetry breaking forces.
    This method opens a new vista of \textit{ab initio} nuclear energy density functionals.
  \end{abstract}
  \maketitle
\end{CJK*}
%%%%%%%%%%%%%%%%%%%%%%%%%%%%%%%%%%%%%%%%%%%%%%%%%% 
% 
% Introduction
% 
\par
\textit{Introduction.}
The energy density functionals (EDFs),
such as Skyrme~\cite{
  Vautherin1972Phys.Rev.C5_626,
  Bender2003Rev.Mod.Phys.75_121},
Gogny~\cite{
  Decharge1980Phys.Rev.C21_1568,
  Berger1991Comput.Phys.Commun.63_365},
the Michigan-three-range-Yukawa type~\cite{
  Bertsch1977Nucl.Phys.A284_399,
  Nakada2003Phys.Rev.C68_014316,
  Nakada2020Int.J.Mod.Phys.E29_1930008} ,
the Fayans ones~\cite{
  Fayans1998JETPLett.68_169,
  Reinhard2017Phys.Rev.C95_064328},
and the covariant ones~\cite{
  Meng2006Prog.Part.Nucl.Phys.57_470, 
  Niksic2011Prog.Part.Nucl.Phys.66_519,   
  Liang2015Phys.Rep.570_1},
successfully reproduce many properties of atomic nuclei in the whole region of the nuclear chart
as well as nuclear matter properties related to the equation of state and neutron stars~\cite{
  Lattimer2012Annu.Rev.Nucl.Part.Sci.62_485,
  Nakatsukasa2016Rev.Mod.Phys.88_045004,
  Roca-Maza2018Prog.Part.Nucl.Phys.101_96,
  Colo2020Adv.Phys.X5_1740061,
  Yoshida2020Phys.Rev.C102_064307}.
\par
The parameter sets of EDFs introduced in the nuclear density functional theory (DFT) are commonly determined phenomenologically 
to reproduce experimental masses and charge radii
as well as nuclear matter properties.
To accomplish a link between microscopic approaches and EDFs,
\textit{ab initio} determination of the parameters of EDFs
as well as its functional form
is highly motivated~\cite{
  Bender2003Rev.Mod.Phys.75_121,
  Stone2007Prog.Part.Nucl.Phys.58_587,
  Kortelainen2010Phys.Rev.C82_024313,
  Stoitsov2010Phys.Rev.C82_054307,
  Drut2010Prog.Part.Nucl.Phys.64_120,
  Bogner2013Comput.Phys.Commun.184_2235,
  Erler2015J.Phys.G42_034026,
  Dobaczewski2016J.Phys.G43_04LT01,
  Shen2019Prog.Part.Nucl.Phys.109_103713,
  Salvioni2020J.Phys.G47_085107,
  Furnstahl2020Eur.Phys.J.A56_85}.
To date, however, a direct correspondence between \textit{ab initio} and DFT remains elusive;
nevertheless, there are different attempts in recent literature to propose a mapping of \textit{ab initio} to DFT,
e.g., in Refs.~\cite{
  NavarroPerez2018Phys.Rev.C97_054304,
  Furnstahl2020Eur.Phys.J.A56_85,
  Salvioni2020J.Phys.G47_085107,
  Zurek2021Phys.Rev.C103_014325,
  Marino2021Phys.Rev.C104_024315}.
Novel methods using functional renormalization group~\cite{
  Polonyi2002Phys.Rev.B66_155113,
  Kemler2013J.Phys.G40_085105,
  Kemler2017J.Phys.G44_015101,
  Liang2018Phys.Lett.B779_436,
  Yokota2019Phys.Rev.C99_024302,
  Yokota2019Prog.Theor.Exp.Phys.2019_011D01,
  Yokota2019Phys.Rev.B99_115106,
  Yokota2021Prog.Theor.Exp.Phys.2021_013A03,
  Yokota2021Phys.Rev.Research3_L012015,
  Yokota:2021klr}
and the inverse Kohn-Sham method~\cite{
  Naito2019J.Phys.B52_245003,
  Accorto2020Phys.Rev.C101_024315,
  Accorto2021Phys.Rev.C103_044304}
have also been proposed,
although these attempts are still limited.
\par
A sophisticated yet practical approach to pin down the EDF parameters is to combine the \textit{ab initio} calculations with phenomenological EDFs.
For instance, the strengths of tensor interaction were determined using a $ G $-matrix calculation~\cite{
  AlexBrown2006Phys.Rev.C74_061303}.
Recently, based on the Skyrme EDF, the strengths of tensor interactions~\cite{
  Shen2019Phys.Rev.C99_034322}
and the charge independence breaking term of the nuclear interaction~\cite{
  Roca-Maza2018Phys.Rev.Lett.120_202501}
have also been determined
by using the Brueckner-Hartree-Fock calculations of proton-neutron drops and symmetric nuclear matter, respectively,
with the bare realistic nuclear interaction.
Although these terms are not included in the original Skyrme EDF,
these attempts and successes are milestones toward constructing \textit{ab initio} EDFs.
Recently, a systematic construction of nuclear EDFs has been proposed, mimicking the 
Jacob's ladder of the Coulomb DFT~\cite{
  Perdew2001AIPConf.Proc.577_1}
and starting from the local density approximation (LDA)~\cite{
  Marino2021Phys.Rev.C104_024315}.
\par
Even though the isospin symmetry breaking (ISB) terms 
are small parts of the nuclear interaction,
effects of the ISB terms for nuclear properties have gotten attention,
for example,
for the isobaric analog states~\cite{
  Suzuki1993Phys.Rev.C47_R1360,
  Colo1998Phys.Rev.C57_3049,
  Kaneko2014Phys.Rev.C89_031302,
  Roca-Maza2018Phys.Rev.Lett.120_202501},
for isobaric multiplet mass equation,
for the Okamoto-Nolen-Schiffer anomaly in the mass differences of mirror nuclei~\cite{
  AlexBrown2000Phys.Lett.B483_49,
  Kaneko2010Phys.Rev.C82_061301,
  Kaneko2013Phys.Rev.Lett.110_172505,
  Baczyk2018Phys.Lett.B778_178,
  Dong2018Phys.Rev.C97_021301,
  Baczyk2019J.Phys.G46_03LT01,
  Dong2019Nucl.Phys.A983_133,
  Hoff2020Nature580_52,
  Baczyk2021Phys.Rev.C103_054320},
and 
for superallowed $ \beta $ decay~\cite{
  Kaneko2017Phys.Lett.B773_521}.
The ISB interaction can be divided into two parts,
the charge symmetry breaking (CSB) and the charge independence breaking (CIB) interactions~\cite{
  Miller2006Annu.Rev.Nucl.Part.Sci.56_253}.
The functional form of those interactions can be derived from effective field theory (EFT) based on the QCD Lagrangian~\cite{
  Miller2006Annu.Rev.Nucl.Part.Sci.56_253}.
The importance of the CSB interaction in the nuclear structure of $ \nuc{Be}{8}{} $
was also discussed by using the Green's function Monte Carlo calculation~\cite{
  Wiringa2013Phys.Rev.C88_044333}.
On the contrary, traditionally, 
the ISB terms have been often neglected in EDF.
We expect that the present investigation will help the nuclear physics and nuclear astrophysics communities to design better functionals in the ISB channels.
It has been also shown that an accurate and detailed theoretical understanding of isospin symmetry breaking in the nuclear medium may entail consequences for the nuclear equation of state, the isobaric analog state, and mass differences of mirror nuclei, among others, e.g., in Refs.~\cite{
  Suzuki1993Phys.Rev.C47_R1360,
  Colo1998Phys.Rev.C57_3049,
  AlexBrown2000Phys.Lett.B483_49,
  Kaneko2010Phys.Rev.C82_061301,
  Kaneko2013Phys.Rev.Lett.110_172505,
  Kaneko2014Phys.Rev.C89_031302,
  Kaneko2017Phys.Lett.B773_521,
  Roca-Maza2018Phys.Rev.Lett.120_202501,
  Baczyk2018Phys.Lett.B778_178,
  Dong2018Phys.Rev.C97_021301,
  Baczyk2019J.Phys.G46_03LT01,
  Dong2019Nucl.Phys.A983_133,
  Hoff2020Nature580_52,
  Baczyk2021Phys.Rev.C103_054320,
  Selva2021Symmetry13_144}.
In nuclear astrophysics, in Ref.~\cite{
  Selva2021Symmetry13_144},
a pioneering study has been published in the context of the mass-radius relation, tidal deformability, and other neutron star properties.
Thus, the study of ISB in EDF is indispensable, and  
the \textit{ab initio} determination of the ISB strength of the EDF is 
desired toward a comprehensive discussion of ISB effects in the whole nuclear chart.
\par
Our motivation in this Letter is to propose a solid methodology to determine the CSB terms
in nuclear EDFs adopting the \textit{ab initio} results.
We add ISB terms in a Skyrme EDF that we choose as an example to show our methodology.
Our conclusions will not depend on that choice as we will discuss later.
The ISB terms are nothing but the lowest-order (local) functional that one could derive in a nuclear effective field theory and corresponds to class II and class III forces in Ref.~\cite{
  Henley1979MesonsinNucleiVolumeI_405}.
Note that 
EFT leading charge-dependent forces have been derived in Ref.~\cite{
  Kolck1995Few-BodySyst.Suppl.9_444},
and 
EFT provides a firm ground why class IV force is small and could be neglected.
Moreover, without explicit pions,
the EFT terms would also reduce what we propose in our Letter.
\par
We will show that the mass difference of mirror nuclei $ \Delta E_{\urm{tot}} $ and the neutron-skin thickness $ \Delta R_{np} $
of doubly-magic nuclei calculated by \textit{ab initio} methods without and with the CSB terms,
once they are available, 
enable us to determine the CSB strength in the functional with
an uncertainty less than $ 6 \, \% $ 
(apart from the uncertainty inherent in the \textit{ab initio} calculations),
independently from other ISB forces,
such as CIB and Coulomb forces.
In the previous studies~\cite{
  Baczyk2018Phys.Lett.B778_178,
  Baczyk2019J.Phys.G46_03LT01},
the mass differences of isobars and isotriplets were adopted to pin down the CIB and CSB strengths 
simultaneously. 
On the other hand, in our approach, we will constrain the CSB interaction separately from the other ISB forces.
In this way, we can set a stricter constraint on the CSB force strength
using the results of \textit{ab initio} calculations.  
% 
% Skyrme EDF
% 
\par
\textit{Proposed CSB and CIB terms in an EDF.}
The isospin symmetry breaking of the nuclear interaction can be divided into two parts;
the CSB interaction $ V_{\urm{CSB}} \equiv V_{nn} - V_{pp} $ 
and the CIB interaction $ V_{\urm{CIB}} \equiv \left( V_{nn} + V_{pp} \right) / 2 - V_{pn}^{T = 1} $,
where $ V_{pp} $, $ V_{nn} $, and $ V_{pn}^{T = 1} $ denote
proton-proton, neutron-neutron, and the $ T = 1 $ channel of proton-neutron nuclear interactions, respectively.
Origins of the CSB interaction are mainly proton-neutron mass difference and $ \pi^0 $-$ \eta $ and $ \rho^0 $-$ \omega $ mixings in the meson exchange process,
and that of the CIB interaction is the mass difference of neutral and charged pions~\cite{
  Coon1982Phys.Rev.C26_2402,
  Miller1990Phys.Rep.194_1,
  Kolck1996Phys.Lett.B371_169,
  Miller1990Phys.Rep.194_1}.
\par
The Skyrme-like zero-range CSB and CIB interactions were introduced in Ref.~\cite{
  Sagawa1995Phys.Lett.B353_7}
as
\begin{subequations}
  \label{eq:int_ISB}
  \begin{align}
    v_{\urm{CSB}} \left( \ve{r}_1, \ve{r}_2 \right)
    & =
      \frac{\tau_{1z} + \tau_{2z}}{4}
      s_0
      \left( 1 + y_0 P_{\sigma} \right)
      \delta \left( \ve{r}_1 - \ve{r}_2 \right), 
      \label{eq:int_CSB} \\
    v_{\urm{CIB}} \left( \ve{r}_1, \ve{r}_2 \right)
    & =
      \frac{\tau_{1z} \tau_{2z}}{2}
      u_0
      \left( 1 + z_0 P_{\sigma} \right)
      \delta \left( \ve{r}_1 - \ve{r}_2 \right), 
      \label{eq:int_CIB}
  \end{align}
\end{subequations}
respectively, where $ \tau_{iz} $ is the $ z $ projection of the isospin operator of nucleon $ i $ ($ i = 1 $, $ 2 $) 
and $ P_{\sigma} = \left( 1 + \ve{\sigma}_1 \cdot \ve{\sigma}_2 \right) / 2 $ is the spin-exchange operator.
This form has been proposed for the sake of simplicity,
whereas momentum-dependent ISB forces have been discussed in
Refs.~\cite{
  Baczyk2019J.Phys.G46_03LT01,
  Baczyk2021Phys.Rev.C103_054320}.
One could write a CIB force
proportional to
the isotensor form $ T_{12} = 3 \tau_{1z} \tau_{2z} - \ve{\tau}_1 \cdot \ve{\tau}_2 $~\cite{
  Wiringa1995Phys.Rev.C51_38}.
However, the physics would remain the same since
the difference between $ T_{12} $ and 
$ \tau_{1z} \tau_{2z} $ is the isoscalar product $ \ve{\tau}_1 \cdot \ve{\tau}_2 $, which 
can be absorbed in the isospin symmetric part of the (bare or effective) interaction.
\par
Accordingly, the CSB and CIB energy densities are derived as~\cite{
  Roca-Maza2018Phys.Rev.Lett.120_202501,
  Naito2020Phys.Rev.C101_064311}
\begin{subequations}
  \label{eq:EDF_ISB}
  \begin{align}
    \ca{E}_{\urm{CSB}} \left[ \rho_p, \rho_n \right]
    & =
      \frac{s_0 \left( 1 - y_0 \right)}{8}
      \left( \rho_n^2 - \rho_p^2 \right),
      \label{eq:EDF_CSB} \\
    \ca{E}_{\urm{CIB}} \left[ \rho_p, \rho_n \right]
    & =
      \frac{u_0}{8}
      \left[
      \left( 1 - z_0 \right) 
      \left( \rho_n^2 + \rho_p^2 \right)
      -
      2 \left( 2 + z_0 \right)
      \rho_n \rho_p
      \right],
      \label{eq:EDF_CIB}
  \end{align}
\end{subequations}
respectively.
For simplicity, in the SAMi-ISB functional~\cite{
  Roca-Maza2018Phys.Rev.Lett.120_202501},
$ y_0 = z_0 = -1 $ 
are chosen to select the spin-singlet channel for both CSB and CIB interactions.
The CIB strength $ u_0 = 25.8 \, \mathrm{MeV} \, \mathrm{fm}^3 $ has been determined by using the Brueckner-Hartree-Fock calculation of symmetric nuclear matter without and with the CIB part of the AV18 bare interaction~\cite{
  Wiringa1995Phys.Rev.C51_38,
  Muether1999Phys.Lett.B445_259,
  Roca-Maza2018Phys.Rev.Lett.120_202501}.
The CSB strength $ s_0 = -26.3 \, \mathrm{MeV} \, \mathrm{fm}^3 $ has been determined to reproduce the experimentally measured isobaric analog energy of $ \nuc{Pb}{208}{} $~\cite{
  Martin2007Nucl.DataSheets108_1583,
  Roca-Maza2018Phys.Rev.Lett.120_202501}.
In this Letter, since $ u_0 $ has been already determined microscopically,
we propose a way to determine $ s_0 $ microscopically.
Since isobaric analog energy is not available from \textit{ab initio} methods at this moment,
we focus on alternative well-established  observables: the nuclear radius and mass,
which are more easily accessible to any \textit{ab initio} method.
% 
% Mechanism
% 
\par
\textit{Physical content of $ \Delta E_{\urm{tot}} $ and $ \Delta R_{np} $.}
First, the reason why $ \Delta E_{\urm{tot}} $ is expected to show a linear dependence on $ s_0 $ can be described by using the energy densities [Eqs.~\eqref{eq:EDF_ISB}].
Approximately, the proton density $ \rho_p $ of $ \nuc{Ca}{48}{} $ is the same as the neutron density $ \rho_n $ of $ \nuc{Ni}{48}{} $ and \textit{vice versa}:
$ \rho_p^{\urm{Ca48}} \left( \ve{r} \right) \simeq \rho_n^{\urm{Ni48}} \left( \ve{r} \right) $
and 
$ \rho_n^{\urm{Ca48}} \left( \ve{r} \right) \simeq \rho_p^{\urm{Ni48}} \left( \ve{r} \right) $.
Using these relationships and Eqs.~\eqref{eq:EDF_ISB}, one finds 
$ \ca{E}_{\urm{CSB}}^{\urm{Ca48}} \simeq - \ca{E}_{\urm{CSB}}^{\urm{Ni48}} $
and 
$ \ca{E}_{\urm{CIB}}^{\urm{Ca48}} \simeq   \ca{E}_{\urm{CIB}}^{\urm{Ni48}} $.
Consequently, the CSB and CIB contributions to $ \Delta E_{\urm{tot}} $ are 
$ \ca{E}_{\urm{CSB}}^{\urm{Ca48}} - \ca{E}_{\urm{CSB}}^{\urm{Ni48}} \simeq 2 \ca{E}_{\urm{CSB}}^{\urm{Ca48}} \sim s_0 $ and
$ \ca{E}_{\urm{CIB}}^{\urm{Ca48}} - \ca{E}_{\urm{CIB}}^{\urm{Ni48}} \simeq 0 $, 
respectively.
Note that deviations from
$ \ca{E}_{\urm{CSB}}^{\urm{Ca48}} - \ca{E}_{\urm{CSB}}^{\urm{Ni48}} = 2 \ca{E}_{\urm{CSB}}^{\urm{Ca48}} $ and
$ \ca{E}_{\urm{CIB}}^{\urm{Ca48}} - \ca{E}_{\urm{CIB}}^{\urm{Ni48}} = 0 $
are due to the Coulomb interaction unless the ISB terms are considered.
\par
The neutron-skin thickness $ \Delta R_{np} $ is also expected to show a dependence on $ s_0 $.
The CSB interaction between a proton and a neutron is exactly zero.
In contrast, the CSB interaction between protons $ v_{\urm{CSB}}^{pp} $ is repulsive,
whereas that between neutrons $ v_{\urm{CSB}}^{nn} $ is attractive,
and $ v_{\urm{CSB}}^{pp} \equiv - v_{\urm{CSB}}^{nn} \sim \left| s_0 \right| $.
As $ \left| s_0 \right| $ become larger, the proton-proton \textit{repulsive} interaction
and the neutron-neutron \textit{attractive} interaction becomes stronger.
Accordingly, the $ \rho_p $ expands and $ \rho_n $ shrinks,
i.e., $ R_p $ becomes larger, and $ R_n $ becomes smaller.
Consequently, $ \Delta R_{np} $ becomes smaller as $ \left| s_0 \right| $ becomes larger.
% 
% Calculation setup
% 
\par
\textit{Calculation setup.}
The nuclear energy density consists of four parts:
the Coulomb term $ \ca{E}_{\urm{Coul}} $,
the isospin symmetric nuclear part $ \ca{E}_{\urm{IS}} $,
the CSB nuclear part $ \ca{E}_{\urm{CSB}} $,
and the CIB nuclear part $ \ca{E}_{\urm{CIB}} $.
The Hartree-Fock-Slater approximation (the Coulomb LDA exchange functional)~\cite{
  Dirac1930Math.Proc.Camb.Philos.Soc.26_376,
  Slater1951Phys.Rev.81_385}
is used for $ \ca{E}_{\urm{Coul}} $~\footnote{
  The treatment of the Coulomb exchange term may cause non-negligible effects for the total energy or the neutron-skin thickness.
  Nevertheless, in this paper, we use the Hartree-Fock-Slater approximation,
  since our purpose in this paper is to discuss the possibility of the \textit{ab initio} determination of the CSB terms.
  Moreover, it will be shown that determination accuracy does not depend on the Coulomb interaction.}.
The SLy4~\cite{
  Chabanat1998Nucl.Phys.A635_231},
SLy5~\cite{
  Chabanat1998Nucl.Phys.A635_231},
SkM*~\cite{
  Bartel1982Nucl.Phys.A386_79},
SAMi~\cite{
  Roca-Maza2012Phys.Rev.C86_031306},
and SAMi-ISB~\cite{
  Roca-Maza2018Phys.Rev.Lett.120_202501}
functionals are used for $ \ca{E}_{\urm{IS}} $
to see whether the choice of functional $ \ca{E}_{\urm{IS}} $ affects the $ s_0 $ dependences of $ \Delta R_{np} $ and $ \Delta E_{\urm{tot}} $.
The parameters of most of the Skyrme functionals, except the SAMi-ISB,
are determined without any ISB effects.
The CIB part $ \ca{E}_{\urm{CIB}} $ is not considered for SLy4, SLy5, SkM*, and SAMi calculations
to keep properties of the original EDFs at most, and to focus only on the CSB effect.
In the calculation with the SAMi-ISB functional,
two types of calculations are performed; 
one is with the CIB functional $ \ca{E}_{\urm{CIB}} $ [Eq.~\eqref{eq:EDF_CIB}]
with the original strength ($ u_0 = 25.8 \, \mathrm{MeV} \, \mathrm{fm}^3 $)
and the other is without the CIB functional.
Hereafter, the former and latter are referred to as the ``SAMi-CIB'' and ``SAMi-noISB'' functionals, respectively.
On top of these calculations, the CSB functional $ \ca{E}_{\urm{CSB}} $ [Eq.~\eqref{eq:EDF_CSB}]
is considered.
By changing gradually the CSB strength from $ - s_0 = 0 $ to $ 50 \, \mathrm{MeV} \, \mathrm{fm}^3 $,
which will be shown by the figures to be a quite reasonable range,
the $ s_0 $ dependences of $ \Delta R_{np} $ and $ \Delta E_{\urm{tot}} $ are discussed.
\par
All the calculations are performed by using a code named \textsc{skyrme\_rpa}~\cite{
  Colo2013Comput.Phys.Commun.184_142}
with a $ 150 \times 0.1 \, \mathrm{fm} $ box.
Spherical symmetry is assumed and the pairing correlations are not considered
since only doubly-magic nuclei are studied.
\par
Before ending this section, 
the difference among SAMi, SAMi-noISB, and SAMi-CIB functionals is explained in detail.
SAMi-noISB and SAMi-CIB functionals share the same $ \ca{E}_{\urm{IS}} $,
whereas $ \ca{E}_{\urm{CIB}} $ for the former and the latter are
$ u_0 = 0 $ and $ 25.8 \, \mathrm{MeV} \, \mathrm{fm}^3 $ in Eq.~\eqref{eq:EDF_CIB}, respectively.
When one applies this method,
first, one constrains the CIB strength $ u_0 $ using nuclear matter 
and, then, constrains the CSB strength $ s_0 $ using the method proposed in this Letter. 
Thus, results of the SAMi-CIB will be referred to.
Nevertheless, to see the effect of the existence of the CIB term, we also show the results with SAMi-noISB.
SAMi functional does not consider the ISB interaction during the fitting
as the other usual Skyrme functionals.
Thus, by comparing the results of SAMi and SAMi-noISB or SAMi-CIB,
one can find the difference between refitting or considering ISB perturbatively.
% 
% Mirror mass difference
% 
\par
\begin{figure}[tb]
  \centering
  \includegraphics[width=1.0\linewidth]{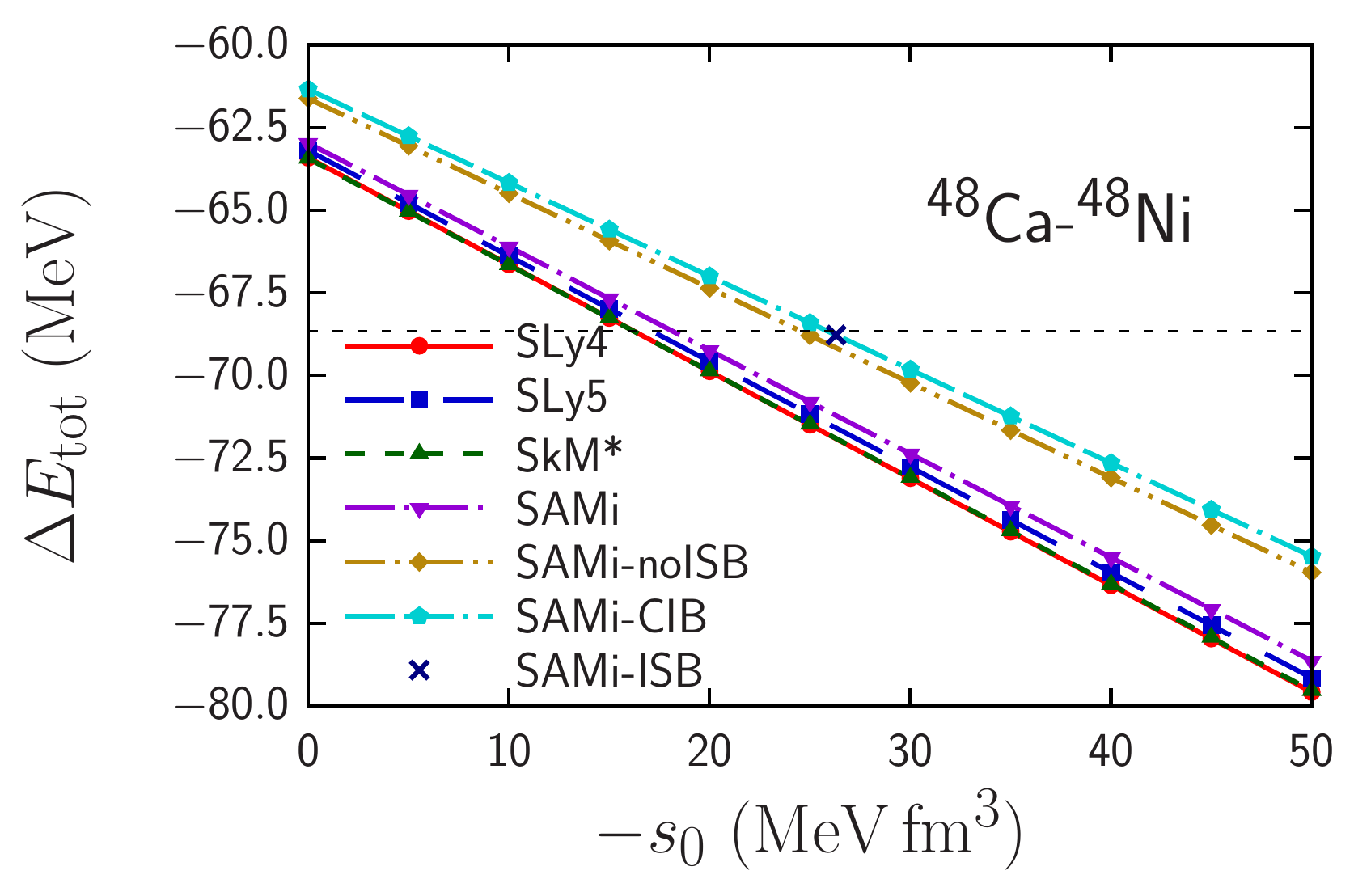}
  \caption{
    Dependence of mass difference of mirror nuclei 
    $ \nuc{Ca}{48}{} $ and $ \nuc{Ni}{48}{} $, $ \Delta E_{\urm{tot}} = E_{\urm{tot}}^{\urm{Ca48}} - E_{\urm{tot}}^{\urm{Ni48}} $, 
    on the CSB strength $ s_0 $.
    For comparison, the result calculated with the original SAMi-ISB functional and experimental data (AME2020) \cite{
      Huang2021Chin.Phys.C45_030002}
    are also shown by the cross point and the horizontal line, respectively.}
  \label{fig:mirror}
\end{figure}

\textit{Mass difference of mirror nuclei.}
Dependence of the mass difference of mirror nuclei $ \nuc{Ca}{48}{} $ and $ \nuc{Ni}{48}{} $,
$ \Delta E_{\urm{tot}} = E_{\urm{tot}}^{\urm{Ca48}} - E_{\urm{tot}}^{\urm{Ni48}} $,
on the CSB strength $ s_0 $ is shown in Fig.~\ref{fig:mirror}.
The circle, square, up-triangle, down-triangle, diamond, and pentagon points represent
the results calculated with the SLy4, SLy5, SkM*, SAMi, SAMi-noISB, and SAMi-CIB functionals, respectively.
For comparison, results calculated with the original SAMi-ISB and experimental data (AME2020)~\cite{
  Huang2021Chin.Phys.C45_030002}
are also shown by the cross point and the horizontal line, respectively.
\begin{table}[tb]
  \centering
  \caption{
    Fitting parameters of $ \Delta E_{\urm{tot}} = a - b s_0 $.}
  \label{tab:mirror}
  \begin{ruledtabular}
    \begin{tabular}{ldd}
      Functional & \multicolumn{1}{c}{$ a $ ($ \mathrm{MeV} $)} & \multicolumn{1}{c}{$ b $ ($ \mathrm{fm}^{-3} $)} \\
      \hline
      SLy4       & -63.4143 & -0.3234 \\
      SLy5       & -63.1975 & -0.3192 \\
      SkM*       & -63.4399 & -0.3220 \\
      SAMi       & -62.9755 & -0.3133 \\
      SAMi-noISB & -61.6251 & -0.2868 \\
      SAMi-CIB   & -61.3474 & -0.2826 \\
    \end{tabular}
  \end{ruledtabular}
\end{table}
\par
One can remarkably find in Fig.~\ref{fig:mirror} that $ \Delta E_{\urm{tot}} $ has a strong linear correlation to the CSB strength $ s_0 $,
and the correlation is universal among the functionals.
Thus, the calculated results are fitted to $ \Delta E_{\urm{tot}} = a - b s_0 $
where fitting parameters determined are shown in Table~\ref{tab:mirror}.
As seen in the figure and table,
$ s_0 $ and $ \Delta E_{\urm{tot}} $ are highly correlated ($ r = 1.000 $),
and
the slope $ b $ is almost universal among Skyrme functional $ \ca{E}_{\urm{IS}} $.
Note that the parameters $ a $ and $ b $ shown in Table~\ref{tab:mirror} are determined within $ 0.5 \, \% $ error.
Accordingly, among these functionals, the slope $ b $ deviates within $ \lesssim 6 \, \% $ around the average value of $ b $ as shown in Table~\ref{tab:nskin_average}.
Thus, once \textit{ab initio} results for $ \Delta E_{\urm{tot}} $ calculated without and with the bare CSB interaction,
$ \Delta E_{\urm{tot}}^{\urm{w/o CSB}} $ and $ \Delta E_{\urm{tot}}^{\urm{w/ CSB}} $,
are obtained, using the averaged value $ \overline{b} $,
we get $ s_0 $ as
$ -s_0 = \left( \Delta E_{\urm{tot}}^{\urm{w/ CSB}} - \Delta E_{\urm{tot}}^{\urm{w/o CSB}} \right) / \overline{b} $.
Since uncertainty of $ \overline{b} $ is less than $ 6 \, \% $, 
the expected uncertainty of $ s_0 $ is also less than $ 6 \, \% $,
assuming the uncertainty associated with the \textit{ab initio} calculation is negligible. 
Note that if the Coulomb interaction is not considered, the parameter $ a $ must be zero.
Thus, $ a $ comes from the Coulomb interaction and its self-consistent effects.
% 
% Neutron-skin thickness
% 
\par
\begin{figure}[tb]
  \centering
  \includegraphics[width=1.0\linewidth]{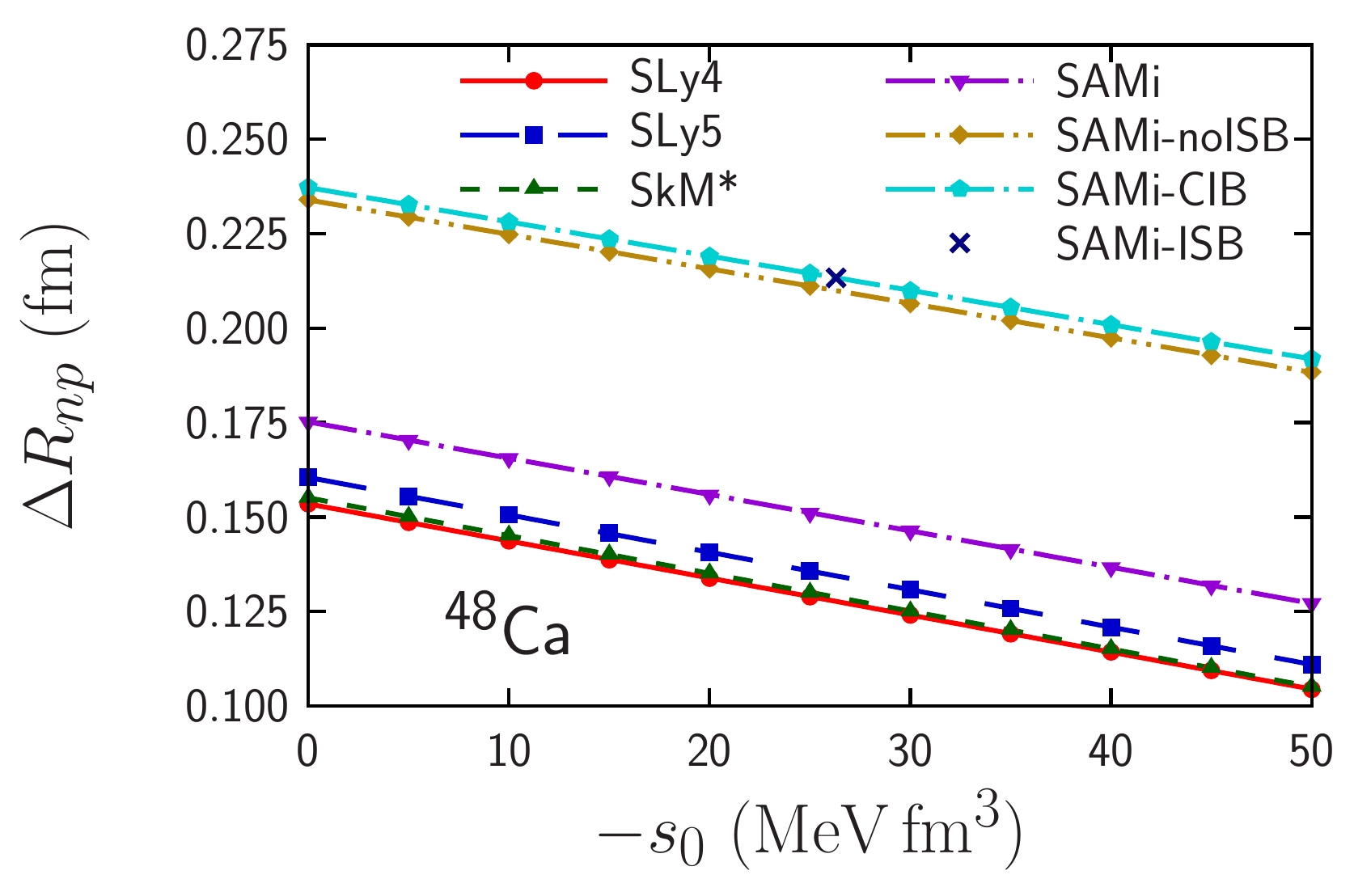}
  \caption{
    Dependence of neutron-skin thickness $ \Delta R_{np} $ of $ \nuc{Ca}{48}{} $
    on the CSB strength $ s_0 $.
    For comparison, the result calculated with the original SAMi-ISB is also shown by the cross point.}
  \label{fig:nskin_020_048}
\end{figure}
\begin{figure}[tb]
  \centering
  \includegraphics[width=1.0\linewidth]{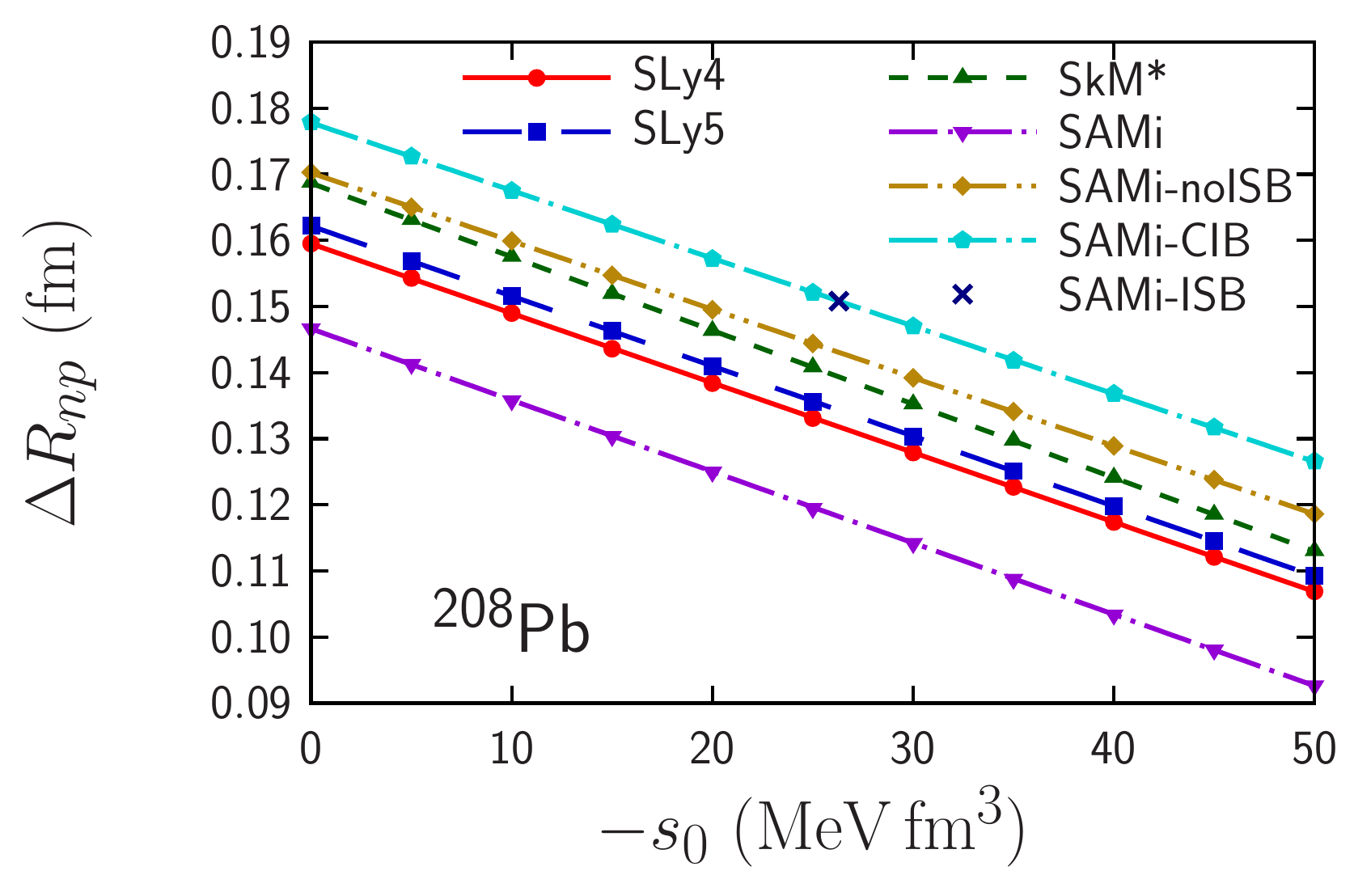}
  \caption{
    Same as Fig.~\ref{fig:nskin_020_048} but for $ \nuc{Pb}{208}{} $.}
  \label{fig:nskin_082_208}
\end{figure}

\textit{Neutron-skin thickness.}
Dependence of the neutron-skin thickness, $ \Delta R_{np} = R_n - R_p $,
of $ \nuc{Ca}{48}{} $ and $ \nuc{Pb}{208}{} $ on the CSB strength $ s_0 $
are shown in Figs.~\ref{fig:nskin_020_048} and \ref{fig:nskin_082_208}, respectively,
where $ R_n $ and $ R_p $ are, respectively, the root-mean-square radii of neutron and proton distributions.
The circle, square, up-triangle, down-triangle, diamond, and pentagon points represent
the results calculated with the SLy4, SLy5, SkM*, SAMi, SAMi-noISB, and SAMi-CIB functionals, respectively.
For comparison, the results calculated with the original SAMi-ISB are also shown by the cross points.
\par
The charge radii of $ \nuc{Ca}{48}{} $ and $ \nuc{Pb}{208}{} $ were measured by electron scattering~\cite{
  DeVries1987At.DataNucl.DataTables36_495}.
The neutron radii of $ \nuc{Ca}{48}{} $ and $ \nuc{Pb}{208}{} $ were measured by the proton scattering~\cite{
  Zenihiro2010Phys.Rev.C82_044611,
  Zenihiro_2018}.
The parity-violating asymmetry,
which is related to the neutron radii
of $ \nuc{Pb}{208}{} $, was also measured by the parity-violating electron scattering
by the PREX and PREX-II collaborations~\cite{
  Abrahamyan2012Phys.Rev.Lett.108_112502,
  Adhikari2021Phys.Rev.Lett.126_172502}
and that of $ \nuc{Ca}{48}{} $ by the CREX collaboration has been measured in the same manner~\cite{
  Horowitz2014Eur.Phys.J.A50_48}.
The calculated lines span a broad interval and
the experimental value obtained by parity-violating electron scattering
($ \Delta R_{np} = 0.283 \pm 0.071 \, \mathrm{fm} $)~\cite{
  Adhikari2021Phys.Rev.Lett.126_172502}
is close to the upper limit of the theoretical interval.
On the other hand, the data by proton scattering
($ \Delta R_{np} = 0.211_{-0.063}^{+0.054} \, \mathrm{fm} $)~\cite{
  Zenihiro2010Phys.Rev.C82_044611}
and the reanalysis of the PREX-II data
($ \Delta R_{np} = 0.19 \pm 0.02 \, \mathrm{fm} $)~\cite{
  Reinhard2021Phys.Rev.Lett.127_232501}
are well compatible with the theoretical predictions.
\begingroup
\squeezetable
\begin{table}[tb]
  \centering
  \caption{
    Averaged values of fitting parameter $ b $
    for $ \Delta E_{\urm{tot}} $ and $ \Delta R_{np} $.
    Standard deviations of $ \overline{b} $ are also shown as $ \Delta \overline{b} $.}
  \label{tab:nskin_average}
  \begin{ruledtabular}
    \begin{tabular}{llddd}
      & & \multicolumn{1}{c}{$ \overline{b} $} & \multicolumn{1}{c}{$ \Delta \overline{b} $} & \multicolumn{1}{c}{$ \Delta \overline{b} / \left| \overline{b} \right| $ ($ \% $)} \\
      \hline
      $ \Delta E_{\urm{tot}} $ & ($ \mathrm{fm}^{-3} $)
        & -0.3079 & 0.0167 & 5.437 \\
      \hline
      $ \Delta R_{np} $ ($ \nuc{Ca}{48}{} $)  & ($ \times 10^{-3} \, \mathrm{MeV}^{-1} \, \mathrm{fm}^4 $)
        & -0.9589 & 0.0364 & 3.795 \\
      $ \Delta R_{np} $ ($ \nuc{Pb}{208}{} $) & ($ \times 10^{-3} \, \mathrm{MeV}^{-1} \, \mathrm{fm}^4 $)
        & -1.0605 & 0.0297 & 2.805 \\
    \end{tabular}
  \end{ruledtabular}
\end{table}
\endgroup
\begingroup
\squeezetable
\begin{table}[tb]
  \centering
  \caption{
    Fitting parameters of $ \Delta R_{np} = a - b s_0 $ for $ \nuc{Ca}{48}{} $ and $ \nuc{Pb}{208}{} $, 
    shown in femtometers for $ a $ and $ \times 10^{-3} \, \mathrm{MeV}^{-1} \, \mathrm{fm}^4 $ for $ b $.}
  \label{tab:nskin}
  \begin{ruledtabular}
    \begin{tabular}{ldddd}
      & \multicolumn{2}{c}{$ \nuc{Ca}{48}{} $} & \multicolumn{2}{c}{$ \nuc{Pb}{208}{} $} \\
      \cline{2-3} \cline{4-5}
      Functional & \multicolumn{1}{c}{$ a $} & \multicolumn{1}{c}{$ b $} & \multicolumn{1}{c}{$ a $} & \multicolumn{1}{c}{$ b $} \\
      \hline
      SLy4       & +0.1535 & -0.9807 & +0.1595 & -1.0525 \\
      SLy5       & +0.1605 & -0.9907 & +0.1622 & -1.0591 \\
      SkM*       & +0.1551 & -1.0002 & +0.1686 & -1.1138 \\
      SAMi       & +0.1752 & -0.9615 & +0.1466 & -1.0800 \\
      SAMi-noISB & +0.2339 & -0.9131 & +0.1702 & -1.0324 \\
      SAMi-CIB   & +0.2372 & -0.9075 & +0.1778 & -1.0255 \\
    \end{tabular}
  \end{ruledtabular}
\end{table}
\endgroup
\par
One can again remarkably find in Figs.~\ref{fig:nskin_020_048} and \ref{fig:nskin_082_208} that $ \Delta R_{np} $ for $ \nuc{Ca}{48}{} $ and $ \nuc{Pb}{208}{} $ have also strong linear correlations to the CSB strength $ s_0 $,
and the correlations are universal among the functionals.
Hence, these calculated results are fitted to $ \Delta R_{np} = a - b s_0 $,
where fitting parameters for $ \nuc{Ca}{48}{} $ and $ \nuc{Pb}{208}{} $ 
are shown in Table~\ref{tab:nskin}.
As in the case of $ \Delta E_{\urm{tot}} $,
$ s_0 $ and $ \Delta R_{np} $ are highly correlated ($ r = 1.000 $),
and
the slope $ b $ is almost universal among Skyrme functionals $ \ca{E}_{\urm{IS}} $.
Note that the parameters $ a $ and $ b $ shown in Table~\ref{tab:nskin} are determined within $ 0.5 \, \% $ error.
Among these functionals, the slope $ b $ deviates within $ \lesssim 4 \, \% $ around the average values of $ b $ as shown in Table~\ref{tab:nskin_average}.
Results for $ \nuc{O}{16}{} $, $ \nuc{Ca}{40}{} $, and $ \nuc{Ni}{48}{} $ are shown in Table \ref{tab:mirror_all} of the Supplemental Material.
Thus, once \textit{ab initio} results of $ \Delta R_{np} $ calculated without and with the bare CSB interaction
are obtained, using the averaged value $ \overline{b} $, we can get $ s_0 $.
Since uncertainty of $ \overline{b} $ for $ \Delta R_{np} $ is also less than $ 4 \, \% $,
the expected uncertainty of $ s_0 $ is also less than $ 4 \, \% $ as in the case of $ \Delta E_{\urm{tot}} $.
\par
The isospin symmetry breaking due to the Coulomb interaction can make a deviation from the linear correlation.
Nevertheless, such an effect is weak enough not to alter the conclusions.
The results are shown in the Supplemental Material.
% 
% 4vs6
% 
\par
\textit{Including or excluding SAMi-ISB functional.}
As seen in Figs.~\ref{fig:mirror}--\ref{fig:nskin_082_208},
the properties of the SAMi-noISB and SAMi-CIB functionals are slightly different
since the ISB parts are already considered in the fit to obtain the parameter set of $ \ca{E}_{\urm{IS}} $.
Due to this difference, the slopes $ b $ of SAMi-noISB and SAMi-CIB are slightly different from the $ b $ of the others.
Consequently, excluding SAMi-noISB and SAMi-CIB functionals in the derivation of $ \overline{b} $ makes slightly smaller deviation $ \Delta \overline{b} $
as shown in the Supplemental Material,
which indicates that the ISB may not be small enough to be treated perturbatively.
Indeed, even if the SAMi-noISB and SAMi-CIB are included,
the deviation $ \Delta \overline{b} $ is small enough.
\par
One can see the effect of refitting by comparing parameters $ b $ derived by SAMi and SAMi-CIB.
Since the fitting criteria for $ \ca{E}_{\urm{IS}} $ for these two functionals are the same,
the difference in $ b $ is only due to the ISB terms.
The strength $ s_0 $ can be determined by
$ \left( \Delta E_{\urm{tot}}^{\urm{w/ CSB}} - \Delta E_{\urm{tot}}^{\urm{w/o CSB}} \right) / b $
or 
$ \left( \Delta R_{np}^{\urm{w/ CSB}} - \Delta R_{np}^{\urm{w/o CSB}} \right) / b $.
Thus, assuming the \textit{ab initio} values for $ \Delta E_{\urm{tot}} $ and $ \Delta R_{np} $ with and without CSB effects are known,
the ratio of $ s_0 $ derived by perturbation ($ s_0^{\urm{perturb}} $) and by refitting ($ s_0^{\urm{refit}} $) reads
$ s_0^{\urm{perturb}} / s_0^{\urm{refit}} = b^{\urm{SAMi-CIB}} / b^{\urm{SAMi}} $,
whose value ranges from $ 0.90 $ (derived by $ \Delta E_{\urm{tot}} $) to $ 0.95 $ (derived by $ \Delta R_{np} $ for $ \nuc{Pb}{208}{} $).
It should be noted that $ b^{\urm{SAMi-CIB}} / b^{\urm{SAMi-noISB}} \approx 0.99 $,
and, thus, fitting for $ s_0 $ before or after the CIB strength $ u_0 $ does not matter.
\par
Recently, the \textit{ab initio} calculations were performed to study the neutron skin of $ \nuc{Ca}{48}{} $ in Ref.~\cite{
  Hagen2016Nat.Phys.12_186}.
Such type of calculations neglecting and including CSB effects will be instrumental to be able to reliably apply the strategy proposed in the present Letter,
especially with the \textit{ab initio} oriented ISB interaction, for example, determined by chiral EFT.
% 
% Conclusion
% 
\par
\textit{Conclusion.}
In this Letter, aiming to construct the \textit{ab initio} EDF, 
we discussed the possibility of determination of the strength of Skyrme-like CSB term $ s_0 $ from the \textit{ab initio} results.
It is found that the mass difference of mirror nuclei $ \Delta E_{\urm{tot}} $ and 
the neutron-skin thickness $ \Delta R_{np} $ of $ \nuc{Ca}{48}{} $ and $ \nuc{Pb}{208}{} $ show a linear dependence on $ s_0 $, 
whereas the CIB term has a much smaller effect on these observables.  
Hence, once \textit{ab initio} calculations of $ \Delta E_{\urm{tot}} $ and $ \Delta R_{np} $ without and with the bare CSB interaction are available,
$ s_0 $ can be determined within $ 6 \, \% $ accuracy by using $ \Delta E_{\urm{tot}} $
and even within $ 4 \, \% $ accuracy by using $ \Delta R_{np} $.
It is important to note that a consistent value for $ s_0 $
using both methods indicates that our assumption in Eq.~\eqref{eq:int_CSB} is reasonable,
whereas a difference among the values of $ s_0 $ derived from $ \Delta E_{\urm{tot}} $ and $ \Delta R_{np} $ may hint to the fact that our ansatz for the CSB functional must be improved. 
The method proposed in this Letter is a feasible way toward \textit{ab initio} nuclear EDFs.
Toward such achievement, future collaborations between \textit{ab initio} and DFT communities are highly desired.
Recently, differences in mass and charge radii of the mirror nuclei and the neutron-skin thickness
have been calculated based on chiral nuclear interactions
with and without the Coulomb interaction
by using the auxiliary field diffusion Monte Carlo and coupled-cluster methods~\cite{
  Novario:2021low}.
Such a type of \textit{ab initio} calculations will likely enable in the future to pin down CSB and CIB in the nuclear medium.
\par
The DFT has been widely applied successfully not only to atomic nuclei but also to atoms, molecules, and solids.
Despite the simpleness of the Coulomb interaction,
an EDF for such systems has been derived without introducing any ansatz 
only in the local density approximation,
and some ansatze should be introduced to derive an EDF with higher-order approximation.
The methodology adopted in this Letter,
i.e., using a quantity sensitive to a specific channel
guided by \textit{ab initio} calculations based on the QCD,
may be applied to other many-body systems.
% 
% Acknowledgement
%
\begin{acknowledgments}
  \par
  T.N. and H.L. thank the RIKEN iTHEMS program
  and the RIKEN Pioneering Project: Evolution of Matter in the Universe.
  T.N. acknowledges the JSPS Grant-in-Aid for JSPS Fellows under Grant No.~19J20543.
  H.L. acknowledges the JSPS Grant-in-Aid for Early-Career Scientists under Grant No.~18K13549
  and
  the Grant-in-Aid for Scientific Research (S) under Grant No.~20H05648.
  H.S. acknowledges the Grant-in-Aid for Scientific Research (C) under Grant No.~19K03858.
  The numerical calculations were performed on cluster computers at the RIKEN iTHEMS program.
\end{acknowledgments}
\end{document}

% 
%%%%%%%%%%%%%%%%%%%%%%%%%%%%%%%%%%%%%%%%%%%%%%%%%% 
\title{
  Supplemental Material for \\
  ``Toward \textit{ab initio} charge symmetry breaking in nuclear energy density functionals''}
\author{Tomoya Naito} % (\CJKfamily{min}{内藤智也})}
\affiliation{
  Department of Physics, Graduate School of Science, The University of Tokyo,
  Tokyo 113-0033, Japan}
\affiliation{
  RIKEN Nishina Center, Wako 351-0198, Japan}
\author{Gianluca Col\`{o}}
\affiliation{
  Dipartimento di Fisica, Universit\`{a} degli Studi di Milano,
  Via Celoria 16, 20133 Milano, Italy}
\affiliation{
  INFN, Sezione di Milano,
  Via Celoria 16, 20133 Milano, Italy}
\author{Haozhao Liang} % (\CJKfamily{gbsn}{梁豪兆})}
\affiliation{
  Department of Physics, Graduate School of Science, The University of Tokyo,
  Tokyo 113-0033, Japan}
\affiliation{
  RIKEN Nishina Center, Wako 351-0198, Japan}
\author{Xavier Roca-Maza}
\affiliation{
  Dipartimento di Fisica, Universit\`{a} degli Studi di Milano,
  Via Celoria 16, 20133 Milano, Italy}
\affiliation{
  INFN, Sezione di Milano,
  Via Celoria 16, 20133 Milano, Italy}
\author{Hiroyuki Sagawa} % (\CJKfamily{min}{佐川弘幸})}
\affiliation{
  Center for Mathematics and Physics, University of Aizu,
  Aizu-Wakamatsu 965-8560, Japan}
\affiliation{
  RIKEN Nishina Center, Wako 351-0198, Japan}
\date{\today}
%%%%%%%%%%%%%%%%%%%%%%%%%%%%%%%%%%%%%%%%%%%%%%%%%% 
\begin{abstract}
  In this supplemental material,
  we show results for $ {}^{16} \mathrm{O} $, $ {}^{40} \mathrm{Ca} $, and $ {}^{48} \mathrm{Ni} $,
  together with those for $ {}^{48} \mathrm{Ca} $ and $ {}^{208} \mathrm{Pb} $.
  We also discuss the effects of the Coulomb interaction.
\end{abstract}
\maketitle
%%%%%%%%%%%%%%%%%%%%%%%%%%%%%%%%%%%%%%%%%%%%%%%%%% 
\renewcommand{\thefigure}{S.\arabic{figure}}
\renewcommand{\thetable}{S.\arabic{table}}
\par
The mass difference of mirror nuclei pair $ \nuc{Ca}{48}{} $ and $ \nuc{Ni}{48}{} $,
$ \Delta E_{\urm{tot}} = E_{\urm{tot}}^{\urm{Ca48}} - E_{\urm{tot}}^{\urm{Ni48}} $,
calculated with the Coulomb interaction
are fitted to $ \Delta E_{\urm{tot}} = a - b s_0 $,
where the fitting parameters  are shown in Table~\ref{tab:mirror_all}.
Meanwhile, the results calculated without the Coulomb interaction are fitted to
$ \Delta E_{\urm{tot}} = - b s_0 $,
whose parameters $ b $ are also shown there.
\par
\begin{figure*}[bp]
  \centering
  \begin{minipage}[t]{0.49\linewidth}
    \centering
    \includegraphics[width=1.0\linewidth]{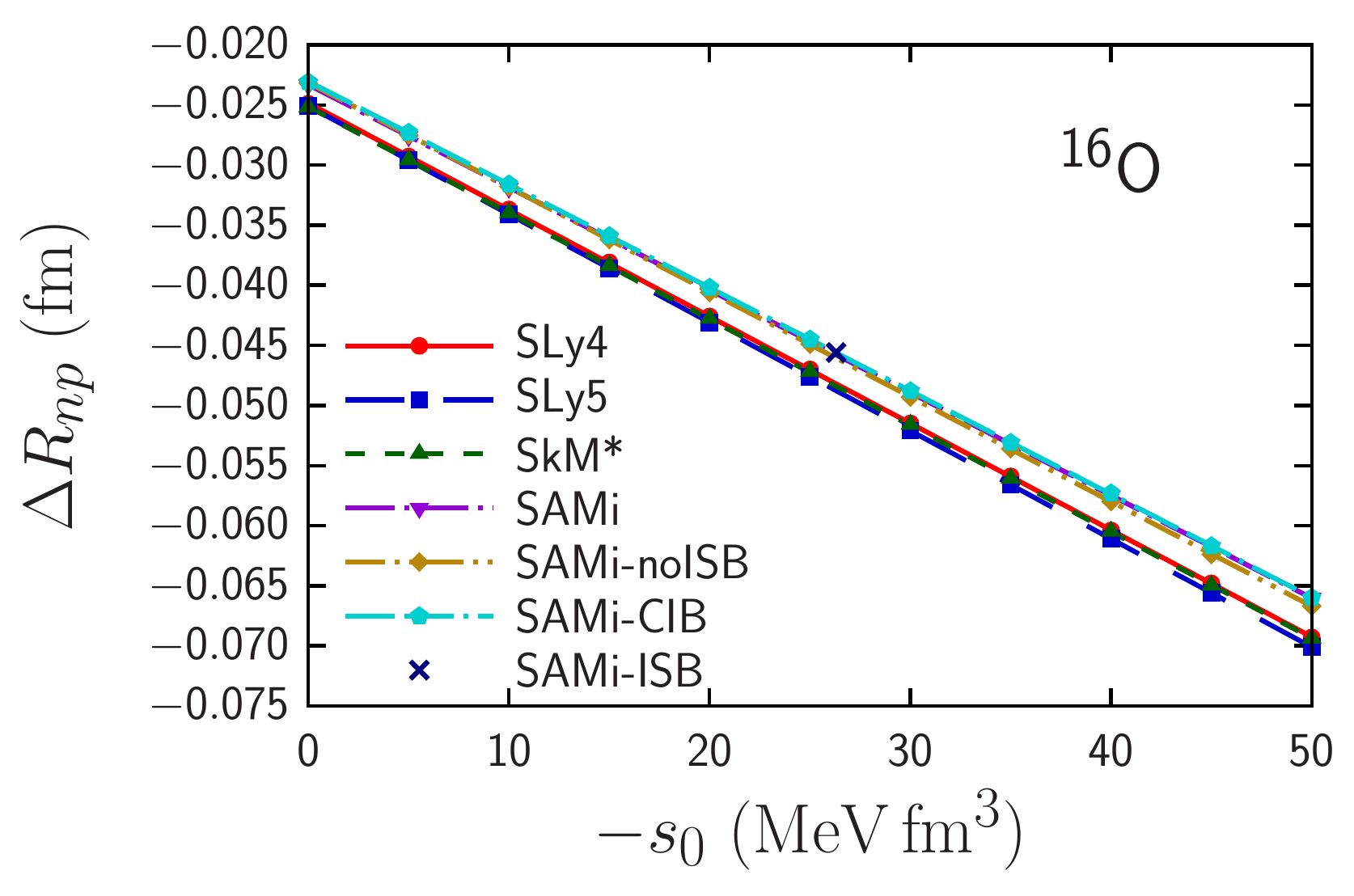}
    \caption{
      Dependence of neutron-skin thickness $ \Delta R_{np} $ of $ \nuc{O}{16}{} $
      on the CSB strength $ s_0 $.
      For comparison, the result calculated with the original SAMi-ISB functional is also shown by the cross point.}
    \label{fig:nskin_008_016}
  \end{minipage}
  \hfill
  \begin{minipage}[t]{0.49\linewidth}
    \centering
    \includegraphics[width=1.0\linewidth]{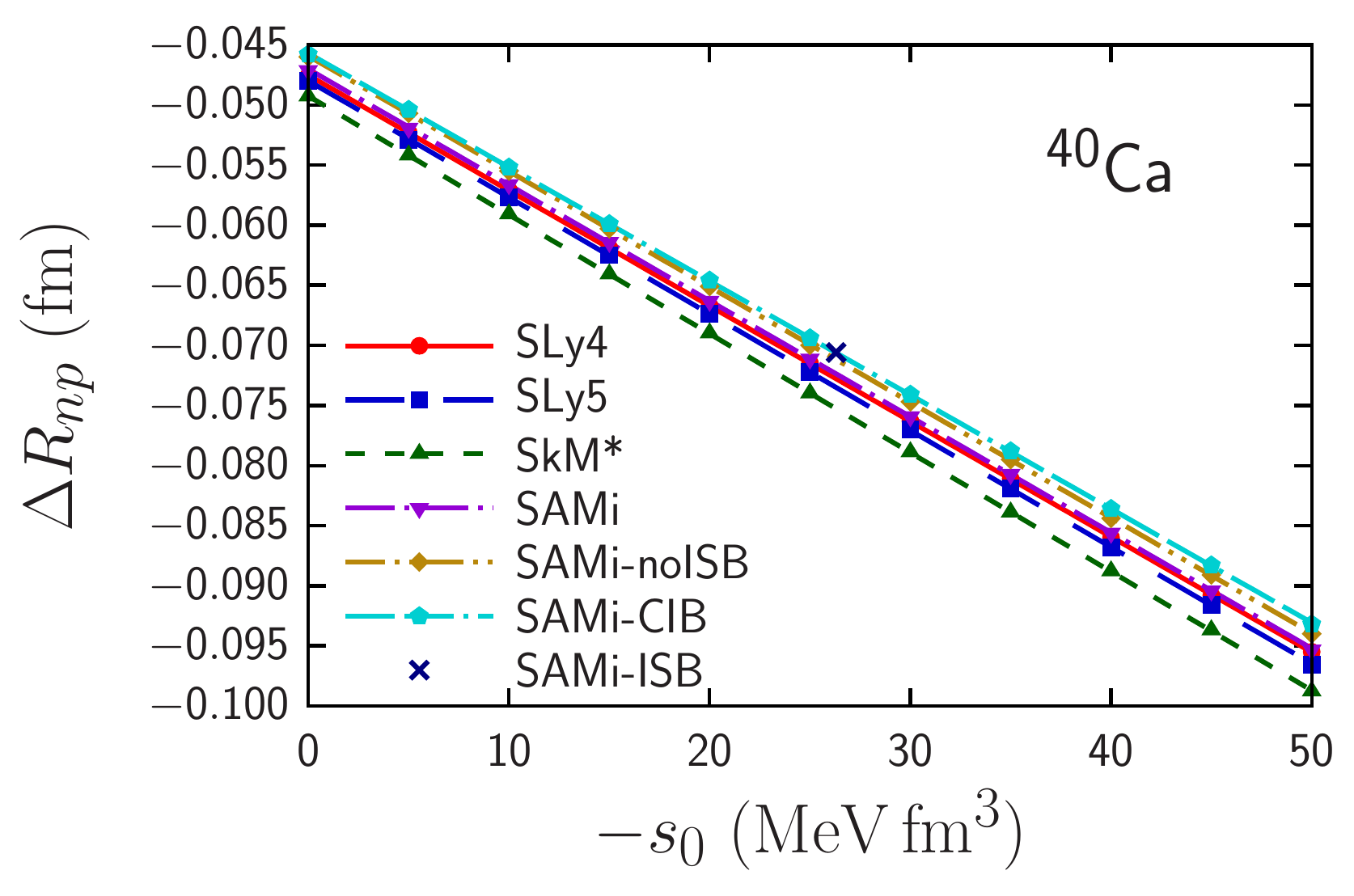}
    \caption{
      Same as Fig.~\ref{fig:nskin_008_016} but for $ \nuc{Ca}{40}{} $.}
    \label{fig:nskin_020_040}
  \end{minipage}
  \begin{minipage}[t]{0.49\linewidth}
    \centering
    \includegraphics[width=1.0\linewidth]{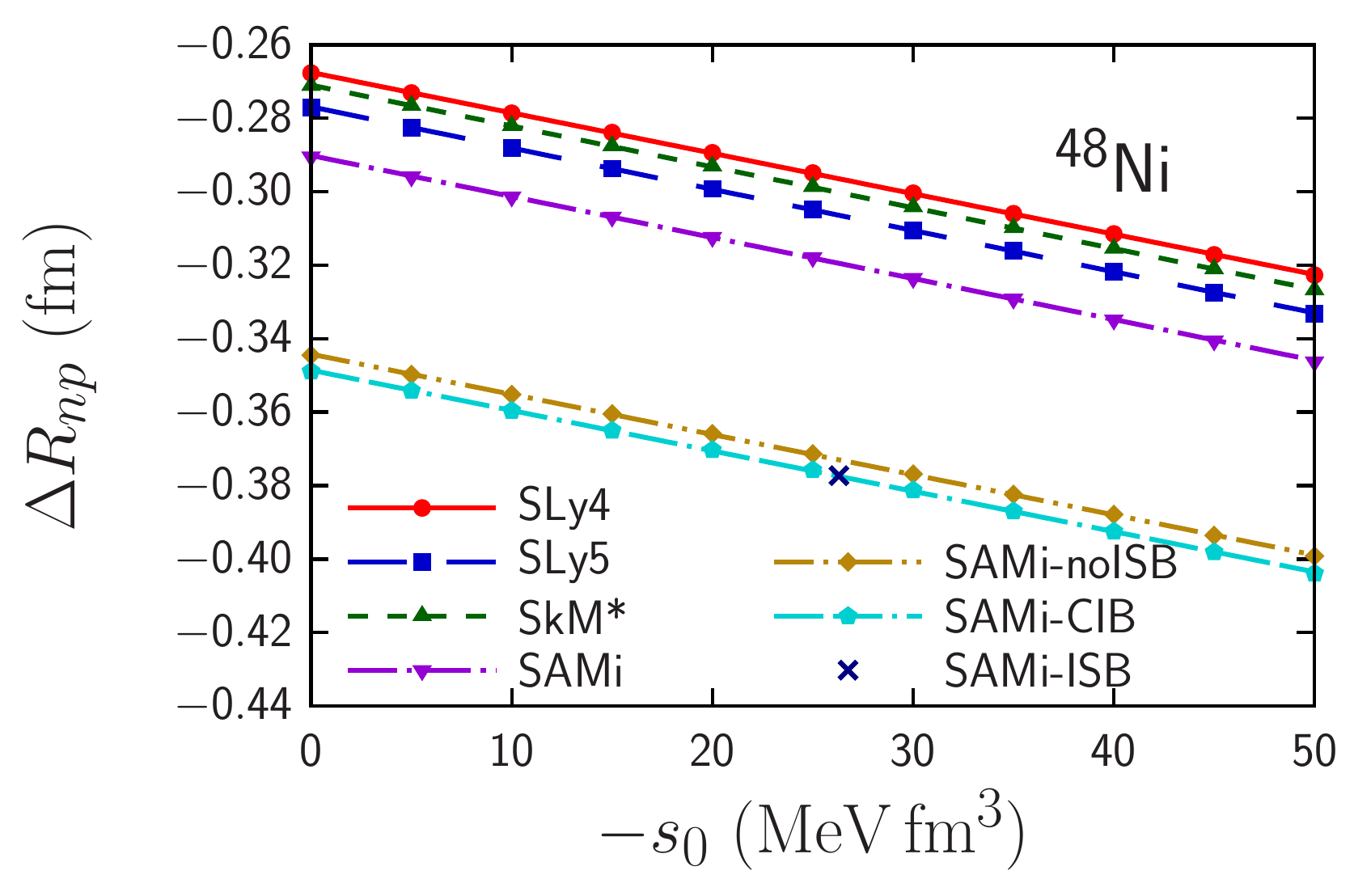}
    \caption{
      Same as Fig.~\ref{fig:nskin_008_016} but for $ \nuc{Ni}{48}{} $.}
    \label{fig:nskin_028_048}
  \end{minipage}
  \hfill
  \begin{minipage}[t]{0.49\linewidth}
  \end{minipage}
\end{figure*}

The CSB strength $ s_0 $ dependences of the neutron-skin thickness $ \Delta R_{np} $ of
$ \nuc{O}{16}{} $, $ \nuc{Ca}{40}{} $, and $ \nuc{Ni}{48}{} $ are shown in
Figs.~\ref{fig:nskin_008_016}, \ref{fig:nskin_020_040}, and \ref{fig:nskin_028_048}, respectively.
These calculated data are fitted to $ \Delta R_{np} = a - b s_0 $,
where the fitting parameters for
$ \nuc{O}{16}{} $, $ \nuc{Ca}{40}{} $, $ \nuc{Ca}{48}{} $, $ \nuc{Ni}{48}{} $, and $ \nuc{Pb}{208}{} $ 
are shown in Table~\ref{tab:nskin_all_CLDA}.
The corresponding results calculated without the Coulomb interaction are fitted to
$ \Delta E_{\urm{tot}} = - b s_0 $ (for $ N = Z $ nuclei)
or 
$ \Delta E_{\urm{tot}} = a - b s_0 $ (for $ N \ne Z $ nuclei).
The fitting parameters are shown in Table~\ref{tab:nskin_all_NoCo}.
Note that all the fittings are highly correlated ($ r = 1.0000 $).
\par
By using $ b $ shown in Tables~\ref{tab:mirror_all} and \ref{tab:nskin_all_CLDA}, 
the average values $ \overline{b} $ and their standard deviations $ \Delta \overline{b} $ are shown in Table~\ref{tab:nskin_average_all}.
Results with all the used functionals, i.e., SLy4, SLy5, SkM*, SAMi, SAMi-noISB, and SAMi-CIB,
are shown in the column entitled ``LDA (All),''
while the results without SAMi-noISB nor SAMi-CIB are shown in the column entitled ``LDA (w/o ISB).''
One can find that the CSB strength $ s_0 $ dependences of $ \Delta R_{np} $ are universal in selected nuclei, regardless of $ N < Z $, $ N = Z $, or $ N > Z $.
Besides, functional independence of the slope $ b $ also holds among the selected nuclei,
and consequently, the standard deviation $ \Delta \overline{b} $ is less than $ 6 \, \% $ of the averaged value $ \overline{b} $,
as discussed in section entitled ``\textit{Including or excluding SAMi-ISB functional}'' in the main text.
\par
The Coulomb interaction also breaks isospin symmetry of the atomic nuclei,
and therefore, $ \Delta E_{\urm{tot}} $ and $ \Delta R_{np} $ becomes finite values even if the ISB terms are not considered.
To see the effect of the Coulomb interaction, fitting parameters $ a $ and $ b $ of $ \Delta R_{np} $ calculated without the Coulomb interaction are shown in Table~\ref{tab:nskin_all_NoCo}.
Note that and $ \Delta R_{np} $ for $ N = Z $ nuclei and $ \Delta E_{\urm{tot}} $ are exactly zero
when neither the CSB nor the Coulomb interaction is switched off.
Hence, these data are fitted with a constraint $ a = 0 $.
The averaged values $ \overline{b} $ and their standard deviation $ \Delta \overline{b} $ are also shown in columns entitled ``No Coulomb (All)'' and ``No Coulomb (w/o ISB)'' of Table~\ref{tab:nskin_average_all}.
It is found that irrespective of the existence of the Coulomb interaction,
the parameter $ b $, and accordingly $ \overline{b} $, hardly changes.
\par
Hence, in short, the discussion in the main text hardly changes with respect to the choice of nuclei or the existence of the Coulomb interaction.
\begin{table}[!bp]
  \centering
  \caption{
    Fitting parameters of $ \Delta E_{\urm{tot}} = a - b s_0 $ for the results with the Coulomb interaction
    and those of $ \Delta E_{\urm{tot}} = - b s_0 $ for the results without the Coulomb interaction.}
  \label{tab:mirror_all}
  \begin{ruledtabular}
    \begin{tabular}{lddd}
      \multicolumn{1}{c}{Functional} & \multicolumn{2}{c}{Coulomb LDA} & \multicolumn{1}{c}{No Coulomb} \\
                                     & \multicolumn{1}{c}{$ a $ ($ \mathrm{MeV} $)} & \multicolumn{1}{c}{$ b $ ($ \mathrm{fm}^{-3} $)} & \multicolumn{1}{c}{$ b $ ($ \mathrm{fm}^{-3} $)} \\ \hline
      SLy4       & -63.4143 & -0.3234 & -0.3440 \\
      SLy5       & -63.1975 & -0.3192 & -0.3397 \\
      SkM*       & -63.4399 & -0.3220 & -0.3435 \\
      SAMi       & -62.9755 & -0.3133 & -0.3325 \\
      SAMi-noISB & -61.6251 & -0.2868 & -0.3039 \\
      SAMi-CIB   & -61.3474 & -0.2826 & -0.2998 \\
    \end{tabular}
  \end{ruledtabular}
\end{table}
\begingroup
\squeezetable
\begin{table*}[!bp]
  \centering
  \caption{
    Fitting parameters of $ \Delta R_{np} = a - b s_0 $ for
    $ \nuc{O}{16}{} $, $ \nuc{Ca}{40}{} $, $ \nuc{Ca}{48}{} $, $ \nuc{Ni}{48}{} $, and $ \nuc{Pb}{208}{} $
    shown in femtometers for $ a $ and $ \times 10^{-3} \, \mathrm{MeV}^{-1} \, \mathrm{fm}^4 $ for $ b $
    calculated with the Coulomb Hartree-Fock-Slater approximation.}
  \label{tab:nskin_all_CLDA}
  \begin{ruledtabular}
    \begin{tabular}{ldddddddddd}
      Functional & \multicolumn{2}{c}{$ \nuc{O}{16}{} $} & \multicolumn{2}{c}{$ \nuc{Ca}{40}{} $} & \multicolumn{2}{c}{$ \nuc{Ca}{48}{} $} & \multicolumn{2}{c}{$ \nuc{Ni}{48}{} $} & \multicolumn{2}{c}{$ \nuc{Pb}{208}{} $} \\
                 & \multicolumn{1}{c}{$ a $} & \multicolumn{1}{c}{$ b $} & \multicolumn{1}{c}{$ a $} & \multicolumn{1}{c}{$ b $} & \multicolumn{1}{c}{$ a $} & \multicolumn{1}{c}{$ b $} & \multicolumn{1}{c}{$ a $} & \multicolumn{1}{c}{$ b $} & \multicolumn{1}{c}{$ a $} & \multicolumn{1}{c}{$ b $} \\
      \hline
      SLy4       & -0.0248 & -0.8884 & -0.0475 & -0.9604 & +0.1535 & -0.9807 & -0.2675 & -1.1005 & +0.1595 & -1.0525 \\ 
      SLy5       & -0.0251 & -0.9000 & -0.0480 & -0.9700 & +0.1605 & -0.9907 & -0.2769 & -1.1225 & +0.1622 & -1.0591 \\
      SkM*       & -0.0251 & -0.8842 & -0.0493 & -0.9893 & +0.1551 & -1.0002 & -0.2710 & -1.1125 & +0.1686 & -1.1138 \\
      SAMi       & -0.0233 & -0.8542 & -0.0470 & -0.9640 & +0.1752 & -0.9615 & -0.2901 & -1.1165 & +0.1466 & -1.0800 \\
      SAMi-noISB & -0.0231 & -0.8713 & -0.0459 & -0.9605 & +0.2339 & -0.9131 & -0.3442 & -1.0938 & +0.1702 & -1.0324 \\
      SAMi-CIB   & -0.0230 & -0.8585 & -0.0457 & -0.9475 & +0.2372 & -0.9075 & -0.3486 & -1.0987 & +0.1778 & -1.0255 \\
    \end{tabular}
  \end{ruledtabular}
\end{table*}
\endgroup
\begingroup
\squeezetable
\begin{table*}[!bp]
  \centering
  \caption{
    Same as Table~\ref{tab:nskin_all_CLDA} but calculated without the Coulomb interaction.
    In $ N = Z $ nuclei, the results are fitted to
    $ \Delta R_{np} = - b s_0 $ instead.}
  \label{tab:nskin_all_NoCo}
  \begin{ruledtabular}
    \begin{tabular}{ldddddddddd}
      Functional & \multicolumn{2}{c}{$ \nuc{O}{16}{} $} & \multicolumn{2}{c}{$ \nuc{Ca}{40}{} $} & \multicolumn{2}{c}{$ \nuc{Ca}{48}{} $} & \multicolumn{2}{c}{$ \nuc{Ni}{48}{} $} & \multicolumn{2}{c}{$ \nuc{Pb}{208}{} $} \\
                 & \multicolumn{1}{c}{$ a $} & \multicolumn{1}{c}{$ b $} & \multicolumn{1}{c}{$ a $} & \multicolumn{1}{c}{$ b $} & \multicolumn{1}{c}{$ a $} & \multicolumn{1}{c}{$ b $} & \multicolumn{1}{c}{$ a $} & \multicolumn{1}{c}{$ b $} & \multicolumn{1}{c}{$ a $} & \multicolumn{1}{c}{$ b $} \\ \hline
      SLy4       & 0 & -0.8823 & 0 & -0.9483 & +0.1917 & -0.9913 & -0.1917 & -1.0071 & +0.2679 & -1.1351 \\
      SLy5       & 0 & -0.8944 & 0 & -0.9594 & +0.1994 & -1.0035 & -0.1994 & -1.0185 & +0.2717 & -1.1411 \\
      SkM*       & 0 & -0.8743 & 0 & -0.9715 & +0.1942 & -1.0102 & -0.1941 & -1.0249 & +0.2839 & -1.2105 \\
      SAMi       & 0 & -0.8438 & 0 & -0.9458 & +0.2141 & -0.9731 & -0.2140 & -0.9956 & +0.2608 & -1.1624 \\
      SAMi-noISB & 0 & -0.8629 & 0 & -0.9454 & +0.2718 & -0.9255 & -0.2717 & -0.9538 & +0.2781 & -1.1145 \\
      SAMi-CIB   & 0 & -0.8498 & 0 & -0.9307 & +0.2747 & -0.9187 & -0.2747 & -0.9469 & +0.2853 & -1.1087 \\
    \end{tabular}
  \end{ruledtabular}
\end{table*}
\endgroup
\begingroup
\squeezetable
\begin{table*}[!bp]
  \centering
  \caption{
    Averaged values of fitting parameter $ b $
    without considering SAMi-noISB nor SAMi-CIB (``w/o ISB'' in title) and
    with all results (``All'' in title).
    Both calculations 
    with the Coulomb LDA exchange functional (``LDA'' in title)
    and without the Coulomb interaction (``No'' in title)
    are shown.
    Standard deviations of $ \overline{b} $ are also shown as $ \Delta \overline{b} $.}
  \label{tab:nskin_average_all}
  \begin{ruledtabular}
    \begin{tabular}{lldddddddd}
      & & \multicolumn{2}{c}{LDA (w/o ISB)} & \multicolumn{2}{c}{LDA (All)} & \multicolumn{2}{c}{No (w/o ISB)} & \multicolumn{2}{c}{No (All)} \\
      & & \multicolumn{1}{c}{$ \overline{b} $} & \multicolumn{1}{c}{$ \Delta \overline{b} $} & \multicolumn{1}{c}{$ \overline{b} $} & \multicolumn{1}{c}{$ \Delta \overline{b} $} & \multicolumn{1}{c}{$ \overline{b} $} & \multicolumn{1}{c}{$ \Delta \overline{b} $} & \multicolumn{1}{c}{$ \overline{b} $} & \multicolumn{1}{c}{$ \Delta \overline{b} $} \\
      \hline
      $ \Delta E_{\urm{tot}} $ & ($ \mathrm{fm}^{-3} $)
        & -0.3195 & 0.0039 & -0.3079 & 0.0167 & -0.3399 & 0.0046 & -0.3272 & 0.0184 \\
      \hline
      $ \Delta R_{np} $ ($ \nuc{O}{16}{} $)   & ($ \times 10^{-3} \, \mathrm{MeV}^{-1} \, \mathrm{fm}^4 $)
        & -0.8817 & 0.0169 & -0.8761 & 0.0163 & -0.8737 & 0.0187 & -0.8679 & 0.0177 \\
      $ \Delta R_{np} $ ($ \nuc{Ca}{40}{} $)  & ($ \times 10^{-3} \, \mathrm{MeV}^{-1} \, \mathrm{fm}^4 $)
        & -0.9709 & 0.0111 & -0.9653 & 0.0127 & -0.9562 & 0.0102 & -0.9502 & 0.0127 \\
      $ \Delta R_{np} $ ($ \nuc{Ca}{48}{} $)  & ($ \times 10^{-3} \, \mathrm{MeV}^{-1} \, \mathrm{fm}^4 $)
        & -0.9833 & 0.0144 & -0.9589 & 0.0364 & -0.9945 & 0.0141 & -0.9704 & 0.0361 \\
      $ \Delta R_{np} $ ($ \nuc{Ni}{48}{} $)  & ($ \times 10^{-3} \, \mathrm{MeV}^{-1} \, \mathrm{fm}^4 $)
        & -1.1130 & 0.0080 & -1.1075 & 0.0104 & -1.0115 & 0.0112 & -0.9912 & 0.0303 \\
      $ \Delta R_{np} $ ($ \nuc{Pb}{208}{} $) & ($ \times 10^{-3} \, \mathrm{MeV}^{-1} \, \mathrm{fm}^4 $)
        & -1.0764 & 0.0239 & -1.0605 & 0.0297 & -1.1623 & 0.0297 & -1.1454 & 0.0340 \\
    \end{tabular}
  \end{ruledtabular}
\end{table*}
\endgroup
% 
%%%%%%%%%%%%%%%%%%%%%%%%%%%%%%%%%%%%%%%%%%%%%%%%%% 
% 